\documentclass[twocolumn,english,prl]{revtex4-2}
\usepackage[T1]{fontenc}
\usepackage[latin9]{inputenc}
\usepackage{color}
\usepackage{babel}
\usepackage{mathtools}
\usepackage{amsmath}
\usepackage{amssymb}
\usepackage{graphicx}
\usepackage[unicode=true,pdfusetitle,
 bookmarks=true,bookmarksnumbered=false,bookmarksopen=false,
 breaklinks=false,pdfborder={0 0 1},backref=false,colorlinks=true]
 {hyperref}
\hypersetup{
 urlcolor={blue}}

\makeatletter
\usepackage{babel}
\usepackage[flushmargin,symbol]{footmisc}

\makeatother

\begin{document}
\title{Hysteretic response to different modes of ramping an external field
in sparse and dense Ising spin glasses}
\author{Mahajabin Rahman and Stefan Boettcher}
\affiliation{Department of Physics, Emory University, Atlanta, GA 30322, USA}
\begin{abstract}
We consider the hysteretic behavior of Ising spin glasses at $T=0$
for various modes of driving. Previous studies mostly focused on an
infinitely slow speed $\dot{H}$ by which the external field $H$
was ramped to trigger avalanches of spin flips by starting
with destabilizing a single spin while few have focused on the effect of different driving methods. First, we show that this conventional protocol imposes a system size dependence. Then, we numerically analyze the response of Ising spin glasses at rates $\dot{H}$ that are fixed as well, to elucidate the differences in the response. Specifically, we compare three different modes of ramping
($\dot{H}=c/N$, $\dot{H}=c/\sqrt{N}$, and $\dot{H}=c$ for constant
$c$) for two types of spin glass systems of size $N$, representing dense networks
by the Sherrington-Kirkpatrick model and sparse networks by the lattice
spin glass in $d=3$ dimensions known as the Edwards Anderson model.
Depending on the mode of ramping, we find that the response of each
system, in form of spin-flip avalanches and other observables, can
vary considerably. In particular, in the $N$-independent mode applied
to the lattice spin glass, which is closest to experimental reality,
we observe a percolation transition with a broad avalanche distribution
between phases of localized and system-spanning responses. We explore implications for combinatorial optimization problems pertaining to
sparse systems.
\end{abstract}
\maketitle

\section{Introduction\label{sec:Introduction}}

Hysteresis is commonly associated with a non-equilibrium,
history-dependent response to a gradually varying external driving
field in a material with bistable behavior often found at sufficiently
low temperatures near a first-order phase transition. There, a barrier
to overcome ordering delays the attainment of lower-energy configurations \citep{bertotti:98,Dahmen1996}. While some aspects are still not fully understood, materials which exhibit hysteretic behavior have important applications as switches or memories, as their non-equilibrium state that the external field drove them into is quite robust against thermal fluctuations and is retained even after that field is turned off \citep{Nan15}. Ideally, they only change state when that field is explicitly reversed. 

There is a long history for the study of hysteresis in ferromagnetic materials \citep{jiles:84}. More pertinent to our purpose here are those
conducted on random field ferromagnets at various levels of disorder,
see the recent work by Spasojevi\'{c} et al \citep{Spasojevic22,Radic21}
and references therein. Experiments on the hysteretic behavior in
spin glasses at low temperature, for example alloys of CuMn, go back at least
to Monod et al. \citep{Monod79}. Bertotti and Pasquale \citep{Bertotti90} have numerically
investigated hysteresis in the Sherrington-Kirkpatrick model (SK)
\citep{Sherrington75} for a range of system sizes and ramping rates
$\dot{H}=dH/dt$ (where we set $dt=1$ to fix the unit of time). Already there,
it was observed that remanent and coercive field both tend to smaller
absolute values for increasing system size. (We find here, in fact, that the hysteresis loop
for SK vanishes in the thermodynamic limit.) Their study of the Barkhausen noise was extended
systematically in the limit of $\dot{H}\to0$ in Ref. \citep{Pazmandi99},
where it was found that the chain reaction of spin flips (termed ''avalanche'') exhibits self-organized critical
(SOC) \citep{BTW,Bak97,BoPa2}. Ref. \citep{Zarand02a} even picked up on the
proposal of Ref. \citep{Bertotti90} to use hysteretic ramping as
a means to find ground states of SK (at $H=0$), an NP-hard problem  \citep{Garey79, Papadimitriou98, Monasson99}, resulting in a heuristic called hysteretic optimization (HO), which consists of demagnetizing disordered models to approximate their ground state.  The critical behavior in SK resulting from hysteretic ramping inspired Ref. \citep{Zarand02a} to liken thermal noise to Barkhausen noise, since both create new spin configurations that can help overcome energetic barriers.  However, while this
HO heuristic proved successful for
SK \citep{Pal06b}, it was found lacking \citep{Goncalves08,Andresen2013}
in producing the prerequisite critical avalanches in sparse
networks of spins, such as for the Edwards-Anderson model (EA), i.e.,
the Ising spin glass on a lattice \citep{Edwards75}. 

Careful consideration of the most marginally stable spins before the
next ramp-up $dH$ in the external field $H$ attributed the critical
behavior to the mutually correlated state these spins attain within
SK \citep{Eastham06,Yan15,Sharma18}. Such correlations are lacking
in EA and other sparse systems in which those marginally stable spins
are likely widely separated in space. 

The properties of marginally
stable spins are distinct also in other manners. For instance, the
typical ramp $dH$ needed to dislodge the next most-unstable spin
in SK is $\sim1/\sqrt{N}$ while it is much smaller for EA, $\sim1/N$, as shown below \citep{Yan15}. Thus, comparing the critical behavior of SK and
lack of critical behavior in EA using inherently different driving modes seems inconsistent. 
Furthermore, as already noted in Ref.~\citep{Spasojevic22}, those
ramps also appear to be unphysical, considering that in a real experiment
one would likely advance the external field via a fixed, constant
$dH$ that destabilizes at once a large (sub-)extensive set of spins.
This is because advancing the field by an increment small enough to trigger exactly
one (or a few) of the spins employed in simulations seems rather difficult
in reality \citep{Zarand02a,Spasojevic22}. Here, we construct three different driving modes, defined by fixed rates unlike quasistatic driving,  and compare their effects on both the EA and SK models to draw a more fair comparison. We find that either $N$-dependent driving mode does not alter the statistics of the SK model, which shows critical behavior regardless. On the contrary, $N$-dependent driving modes on the EA generate exponential avalanche distributions, where ramping the magnetic field creates a broader distribution of avalanches. Lastly, we identify the rate at which there is a percolation threshold in the EA model and evaluate its meaningfulness in the context of problems for which the computing time grows exponentially with complexity, without a proven polynomial-time solution (i.e NP-hard problems) \citep{Garey79, Papadimitriou98, Monasson99}.

This paper is organized as follows: Section \ref{sec:Spin-Glass-Models} describes the two models we used (SK and EA) and three distinct hysteretic driving protocols they were both subject to. Section \ref{sec:Results} discusses first the asymptotic properties of the hysteresis loop for both EA and SK to investigate the dependence of hysteresis on the systems' internal degrees of freedom. It then analyses the behavior for EA in terms of percolation. We also consider the relation between the percolation of spin flips and their ability to approximate ground state energies. In Section \ref{Conclusions}, we conclude with a summary and a discussion of our findings. 

\section{Models and Methods\label{sec:Spin-Glass-Models}}

We study spin glasses with the Hamiltonian 
\begin{equation}
{\cal H}_{SG}=-\frac{1}{2}\sum_{i=1}^{N}\sum_{j\in\mathcal{N}_{i}}J_{ij}s_{i}s_{j}-H_{\text{ext}}\,\sum_{i=1}^{N}s_{i},
\label{eq:SG-Hamiltonian}
\end{equation}
where ${\cal N}_{i}$ denotes the neighborhood of site $i$. We obtain the Edwards-Anderson (EA) model \citep{Edwards75} when spins are located on a $d$-dimensional hyper-cubic lattice, where each spin $i$ is connected only to the $2d$ other spins in its neighborhood ${\cal N}_{i}$.  In the mean-field Sherrington-Kirkpatrick
(SK) model \citep{Sherrington75}, the neighborhood ${\cal N}_{i}$ consists of all $N-1$ other spins (aside from $i$ itself). Thus, there is a fundamental difference between the two models in that the SK has no locality, whereas EA is a local system. The couplings $J_{ij}$ are assigned
from a Gaussian distribution with variance $\left\langle J^{2}\right\rangle =\frac{1}{N}$
for the SK model, and $\left\langle J^{2}\right\rangle =\frac{1}{2d}$
for the EA model. In the following, we only consider the cubic case
($d=3$) for EA. 

The stability $\lambda_{i}$ of each spin $s_{i}$ takes into account its coupling
with the global external field $H_{\text{ext}}$ and with the local
field imposed by its neighbors $s_{j}$ through their mutual bond $J_{ij}$: 
\begin{equation}
\lambda_{i}=s_{i}\left(\sum_{j\in\mathcal{N}_{i}}J_{ij}s_{j}+H_{\text{ext}}\right).
\label{eq:lambda_stability}
\end{equation}

Studying Ising spin glasses along a hysteresis loop at zero temperature
involves slowly ramping an external magnetic field $H_{\text{ext}}$. It has become
conventional to ramp $H_{\text{ext}}$ only as much as is needed for one spin $s_i$ to be
destabilized  ($\lambda_{i}<0$ for only one site $i$) \citep{bertotti:98,Pazmandi99,Zarand02a,Pal06b,Goncalves08,Andresen2013,Yan15}. 
However, other modes of ramping are conceivable.
The process of relaxing the system following an update of the external
field will then involve flipping a causal sequence of destabilized
spins, thus creating an avalanche that lasts until all spins are again
stable ($\lambda_{i}\geq0$ for all $i$). This relaxation  and destabilization
protocol then repeats until magnetic saturation is reached, unless
some other termination condition is specified. 

\begin{figure}
\hfill{}\includegraphics[viewport=0bp 80bp 630bp 530bp,clip,width=1\columnwidth]{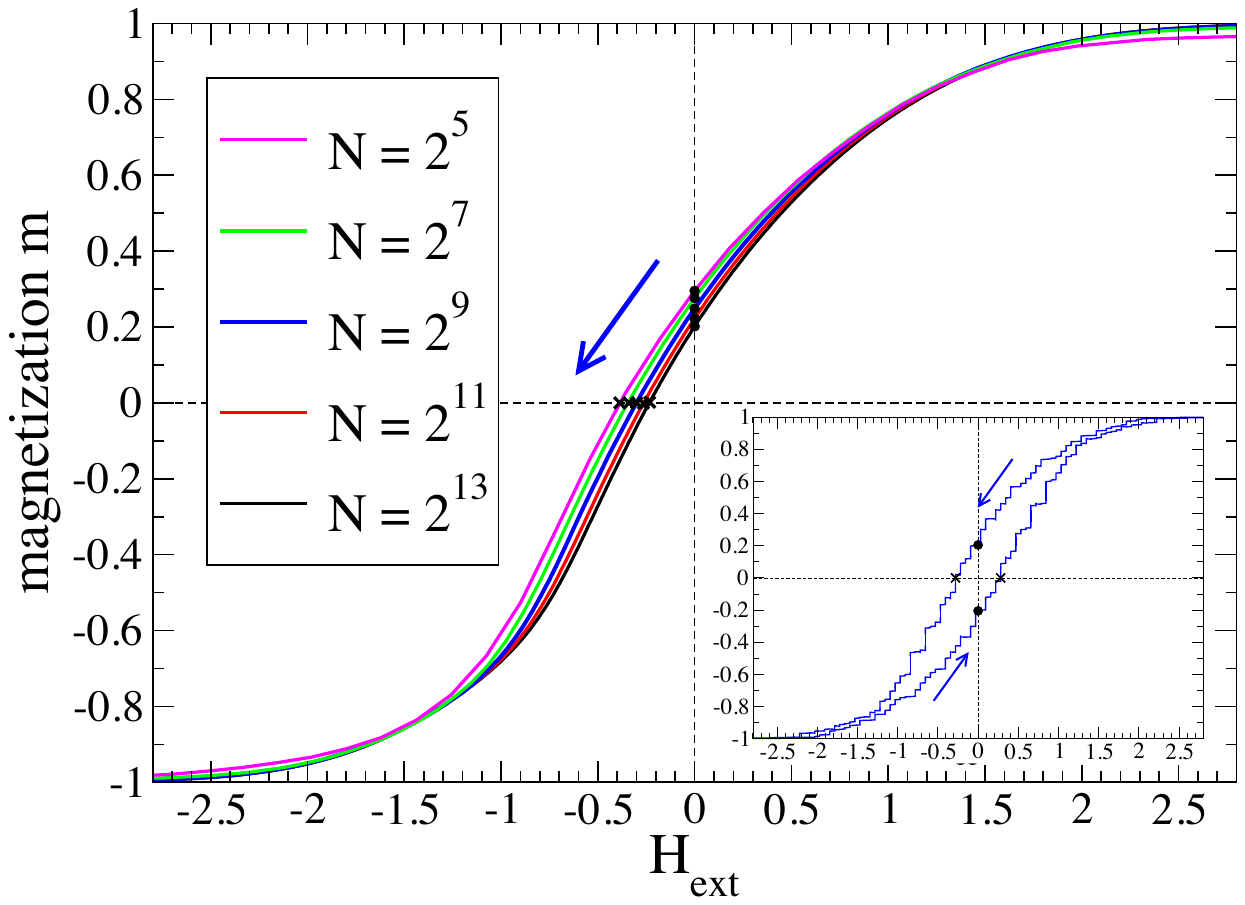}\hfill{}
\caption{\label{fig:SKhyster}Hysteretic behavior in SK at $T=0$, showing
the magnetization per spin $m$ attained as a function of the external
field $H_{{\rm ext}}$ reached by either decreasing (upper branch,
down-arrow) or increasing (lower branch, up-arrow) the field by an
increment $dH$ sufficient to destabilize the next least-stable spin.
As illustrated in the inset for a single run on an instance with $N=2^{10}$,
such a protocol results in a stair-casing behavior of $m$ with horizontal
plateaus due to finite jumps $dH(\sim1/\sqrt{N})$ in $H_{{\rm ext}}$,
followed by instantaneous avalanches of spin-flips leading to vertical
drops in $m$ until all spins are again stabilized at this value of
the field. In the main panel, we have averaged $m$ for the same protocol
over $10^{5}$ runs using distinct disorder instances at the given
system sizes (with only the upper branches being shown). The points
of remanent magnetization $m_{0}$ at $H_{{\rm ext}}=0$ are marked
by black circles, and those of the coersive fields $H_{{\rm ext}}^{{\rm coer}}$
when $m=0$ by black crosses. By both measures, the hysteretic loop
appears to shrink for increasing $N$, see Fig.~\ref{fig:SKhysterExtra}. }
\end{figure}

Fig.~\ref{fig:SKhyster} shows the typical resulting hysteretic behavior
by example of SK averaged over many runs of the conventional protocol and over a range of different
system sizes, albeit for only the upper half of the loop. The inset
illustrates the full hysteresis loop for a single instance, showing
the discrete steps at the interplay of a destabilizing ramp $dH$
and subsequent relaxation behavior. Starting from a fully magnetized
state with magnetization per spin $m=+1$, say, at large positive
field ($H_{{\rm ext}}\gtrapprox2.5$ here), $H_{{\rm ext}}$ is decreased
in increments $dH$ until all spins are aligned downward ($m=-1$)
with a large negative field ($H_{{\rm ext}}\lessapprox-2.5$), from
where the process can be reversed with similar discrete \emph{increases} in
$H_{{\rm ext}}$. The hysteretic process is marked by a non-zero remanent
magnetization $m_{0}$ at the point when the external field passes
$H_{{\rm ext}}=0$ as well as a non-zero coercive field $H_{{\rm ext}}^{{\rm coer}}<0$
needed to bring the magnetization $m$ to vanish. Both are signs of
the non-equilibrium nature of the process, since the equilibrium behavior
of these spin glasses at $T=0$ would have $m=0$ when cooled at $H_{{\rm ext}}=0$. 

\begin{figure}
\hfill{}\includegraphics[viewport=0bp 80bp 630bp 530bp,clip,width=1\columnwidth]{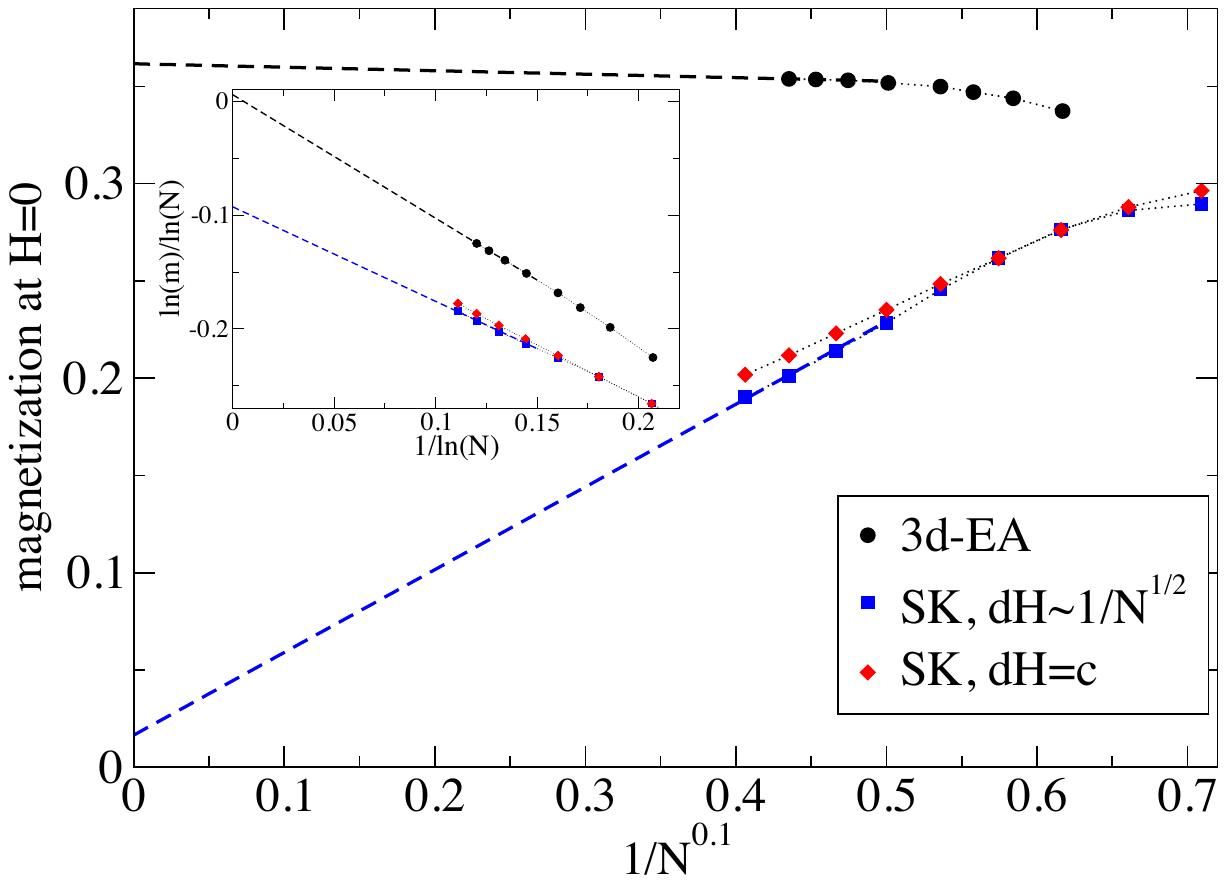}\hfill{}
\caption{\label{fig:SKhysterExtra}Extrapolation plot for the average magnetization
per spin $m_{0}$ attained at $H_{{\rm ext}}=0$ in the hysteresis
loop for increasing system sizes $N$ of SK (as shown in Fig.~\ref{fig:SKhyster})
and EA. The inset extrapolates for the exponent $\omega\sim\log m_{0}/\log N$
in the scaling Ansatz $m_{0}\sim A/N^{\omega}$, suggesting that $\omega\approx0.1$
for SK (blue dashed line) and $\omega\approx0$ for EA (black dashed
line) in the thermodynamic limit (i.e., for $1/\log N\to0$). These
values for $\omega$ are validated in the main panel, where the same
data for $m_{0}$ is plotted as function of $1/N^{0.1}$, extrapolating
about linearly to zero for SK for $N\to\infty$, while for EA $m_{0}$
remains finite. For SK, there is almost no difference in the result
for fixed $dH=c$ or for $dH\sim1/\sqrt{N}$ with $N=2^{5,\ldots,13}$.
For EA, $dH\sim1/N$ with $N=L^{3}$ and $L=5,6,7,8,10,\ldots,16$.}
\end{figure}

The above mentioned conventional protocol is somewhat deterministic
and has implicit $N$ dependence. This dependence arises because destabilization
involves only the weakest spin. In a stable configuration, it is $\lambda_{i}>0$
for all $i$ in Eq.~(\ref{eq:lambda_stability}), with the most unstable
spins being the smallest. These populate the low end, $\lambda\to0$,
of the distribution of all stabilities, to wit, $P(\lambda)\sim\lambda^{\theta}$
for some $\theta>-1$. Then, the fraction of spins with stability
$\lambda_{i}<\lambda$ is given by $n(\lambda)/N=\int_{0}^{\lambda}d\lambda^{\prime}P(\lambda^{\prime})\sim\lambda^{\theta+1}$.
Thus, the typical spacing in stability between the weakest spins is
given by $d\lambda\sim N^{-\frac{1}{1+\theta}}$ and, accordingly,
to dislodge just one (or a finite number) of the spins, we need to ramp
the field by $dH\sim N^{-\frac{1}{1+\theta}}$ \citep{Yan15}. For
SK, it is well-known that $\theta=\frac{1}{2}$ \citep{Bray81,Eastham06,Boettcher07b},
meaning that $dH\sim\frac{1}{\sqrt{N}}$ on average for the conventional
protocol. For EA it is $\theta=0$ \citep{Boettcher07b}, which makes
the difference between consecutive weakest spins, and thus the ramp
needed to dislodge a finite number of them in the conventional
protocol, scale as $dH\sim\frac{1}{N}$ instead. For any realistic driving mechanism
applied in an experiment, it stands to reason that $dH$ would be
a constant value independent of $N$ \citep{Zarand02a,Spasojevic22}.
Hence, we explore the following three driving modes: (1) $dH=c/N$,
(2) $dH=c/\sqrt{N}$, and (3) $dH=c$. For modes (1) and (2), we set $c = 1$. For mode (3), we try a range of $c$ values from $0.05-0.80$. Note that although we maintain a fixed value $c=1$ for each update with methods (1) and (2), where $dH$ scales with some power of $N$, our results are quite representative of averaged measurements, $\left\langle dH\right\rangle \sim1/N$ for SK
or $\left\langle dH\right\rangle \sim1/\sqrt{N}$ for EA, obtained with the conventional, variable driving \citep{Pazmandi99,Goncalves08,Andresen2013,Yan15}.
Further, if we drive SK with method (1), almost always no weak spin
is in reach of a fixed change by $dH$, resulting in many empty avalanches, those lacking even an initial spin flip, 
since typically $\sim\sqrt{N}$ updates are needed to dislodge
the next weakest spin. Thus, aside from those empty avalanches, driving
method (1) for SK will not be different from method (2). Only if we
drive EA with method (2), or EA and SK with method (3), do we expect
new results different from those earlier studies.

\begin{figure*}
\hfill{}\includegraphics[width=150mm]{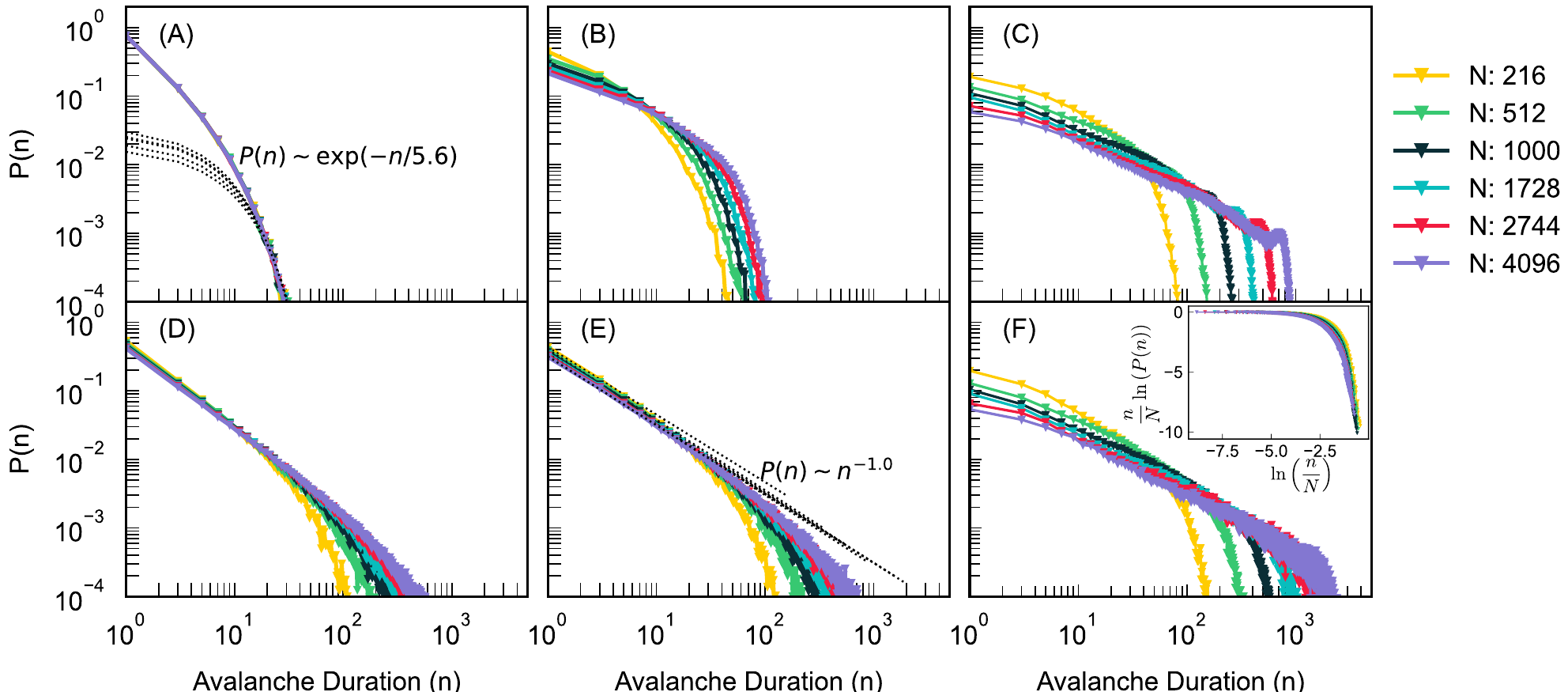}\hfill{}
\caption{\label{fig:n-dependent}The top and bottom rows show distributions
$P(n)$ for avalanche durations $n$ along the hysteresis loop of
the EA and SK model, respectively. (A) and (D) show the
resulting $P(n)$ for a ramping rate of $dH=1/N$ and (B) and (E)
for $dH=1/\sqrt{N}$, for a range of system sizes $N$. In the third column, (C) and (F) show $P(n)$ for $dH$, or $c=0.25$ for various system sizes. In the inset of (F), the distributions are collapsed by rescaling the duration ($n$) and $P (n)$ based on their $N$ dependence. Ignoring empty avalanches, (D) and (E) for
SK are indistinguishable, each showing power-law decay and the size-dependent
scaling in the cut-off characteristic of SOC, which is absent in (A)
and (B) for EA with an exponentially decreasing $P(n)$ where only
the sizeable number of uncorrelated spins triggered ($\sim\sqrt{N}$)
at each ramp $dH$ affects a perceptible shift. For the size-independent
ramp $dH=c$, broader avalanche durations arise in both models.
}
\end{figure*}

\section{Results\label{sec:Results}}

\subsection{Hysteresis Loop in SK and EA\label{subsec:Hysteresis-Loop}}

Before we consider details of the avalanche statistics, we explore
some noteworthy asymptotic properties of the hysteresis loop for EA
and SK. A hysteresis effect technically exists only in the thermodynamic
limit ($N\to\infty$), where any experiment would be conducted. However,
by that standard, SK does not appear to have any measurable hysteresis
loop, although avalanches would continue to exist, independent of
which driving method is employed. Measuring, for example, the remanent
magnetization $m_{0}$ at $H_{{\rm ext}}=0$ we find that it saturates
at a finite value for EA while it seems to vanish slowly with increasing
$N$. Thus, the open loop, as seen for instance in the inset of Fig.~\ref{fig:SKhyster},
appears to be only a transient feature in SK. In turn, the corresponding
loop in EA remains quite stable. As shown in Fig.~\ref{fig:SKhysterExtra},
$m_{0}$ appears to vanish for the simulations of SK (as suggested
by Fig.~\ref{fig:SKhyster}), unlike for EA. In the inset, we
attempt to determine the rate at which $m_{0}$ reaches its thermodynamic
limit, assuming
\begin{equation}
\left\langle m_{0}\right\rangle \sim\frac{A}{N^{\omega}},
\label{eq:m0extra}
\end{equation}
which is purely empirical.
Then, plotting $\log\left\langle m_{0}\right\rangle /\log N\sim-\omega+\log A/\log N$
as a function of $1/\log N$ provides the value of $\omega$ at the
intercept ($N=\infty$). While the data for EA clearly implies that
$\omega=0$, it suggests $\omega\approx0.1$, which is corroborated
in the main panel of Fig.~\ref{fig:SKhysterExtra} by the near-linear
decay of $m_{0}$ with $1/N^{0.1}$ (although $1/\log N$ 
would also be conceivable).

\subsection{Avalanche Statistics during Hysteresis\label{subsec:Avalanche-Statistics}}

In an all-to-all connected model such as the SK, it is likely that,
given enough mutual frustration \citep{Yan15}, even the weakest spin
may launch an avalanche of spin flips. However, in sparse systems where
the weakest spin may be connected only to its closest neighbors, the
system will relax rather quickly due to a lack of long-range correlations
in frustration. Accordingly, it has been found that SK achieves a
SOC state \citep{Pazmandi99}, whereas EA exhibits only relatively
brief avalanches \citep{Goncalves08,Andresen2013}.

To characterize avalanches, we used the number of spin flips as a measure of the avalanche duration $n$ within an increment $dH$ along the hysteresis loop. We collected avalanche durations $n$ at every increment $dH$ along the hysteresis loop, repeating for over 100 loops. From these runs, the $n$ values were linearly binned, and the frequency of points within each bin were counted and normalized to form the probability distribution $P(n)$. 
Figure \ref{fig:n-dependent} shows the  probability distribution $P(n)$ of $n$ for the cubic EA lattice and the SK model.  
For the conventional driving mode, represented by (1) for EA and (2) for SK, the dependence
of the ramping rate $dH$ on $N$ ends up imposing an unequal comparison
between the two models. Thus, we force the driving modes to be the
same to create a side-by-side comparison. In Fig.~\ref{fig:n-dependent},
from left to right, panels in each row correspond to modes (1), (2),
and (3), where the top row refers to EA and the bottom to
SK. As Ref. \citep{Yan15} noted, SK displays critical behavior
all along the hysteresis loop, which is why there is consistently
a power law distribution of the avalanches irrespective of the actual
driving. Even when driven in mode (3), a power-law distribution of
avalanche durations persist, see panel (F), where the distributions of the different system sizes can easily be collapsed.

\begin{figure*}
\hfill{}\includegraphics[width=0.95\textwidth]{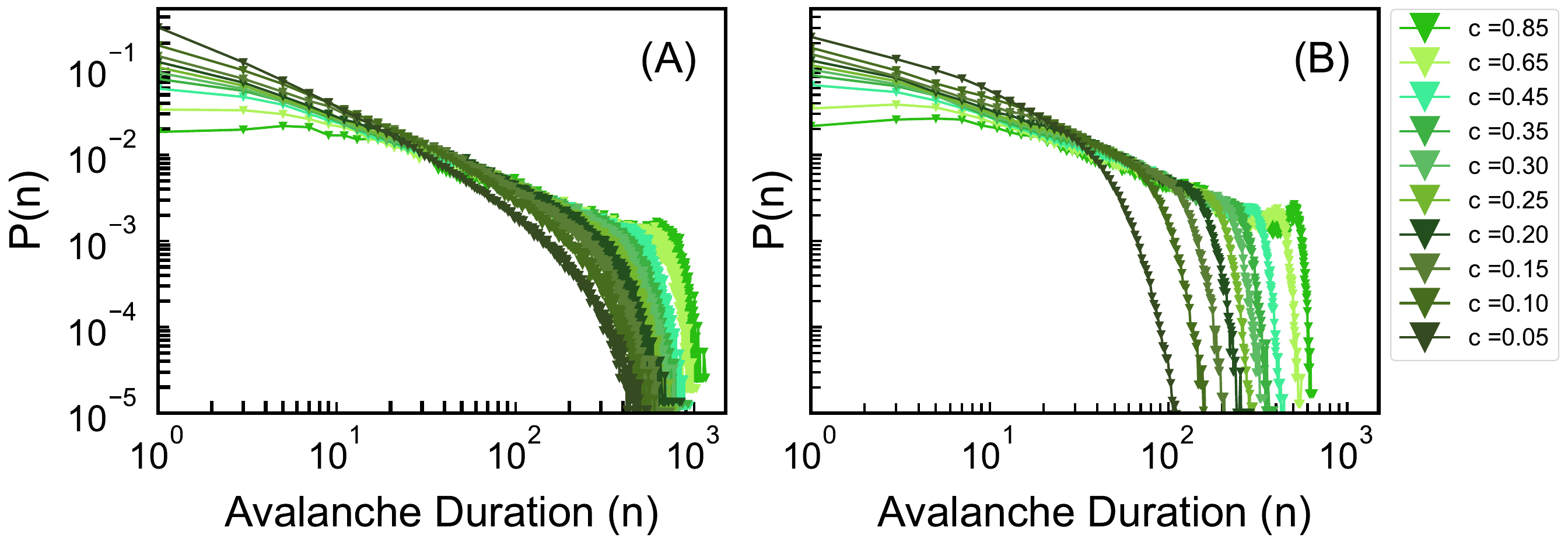}\hfill{}
\caption{\label{fig:c-dependent} Probability $P(n)$ of avalanche durations $n$, with the fixed system size $N = 1000$, and various values of $dH=c$. (A) shows $P(n)$ for SK, and (B) for EA. With increasing values of $c$, the distributions do become broader for both models. However, the behavior near the cut-off begins to change once $c>0.45$, where a second peak emerges in lieu of the exponential fall off seen in Figure \ref{fig:n-dependent} (F). The increasing number of dislodged spins account for this peak. Their non-negligible contribution indicates that they no longer trigger a large enough cascade for the system to be critical.}
\end{figure*}

\begin{figure}
\hfill{}\includegraphics[width=0.9\columnwidth]{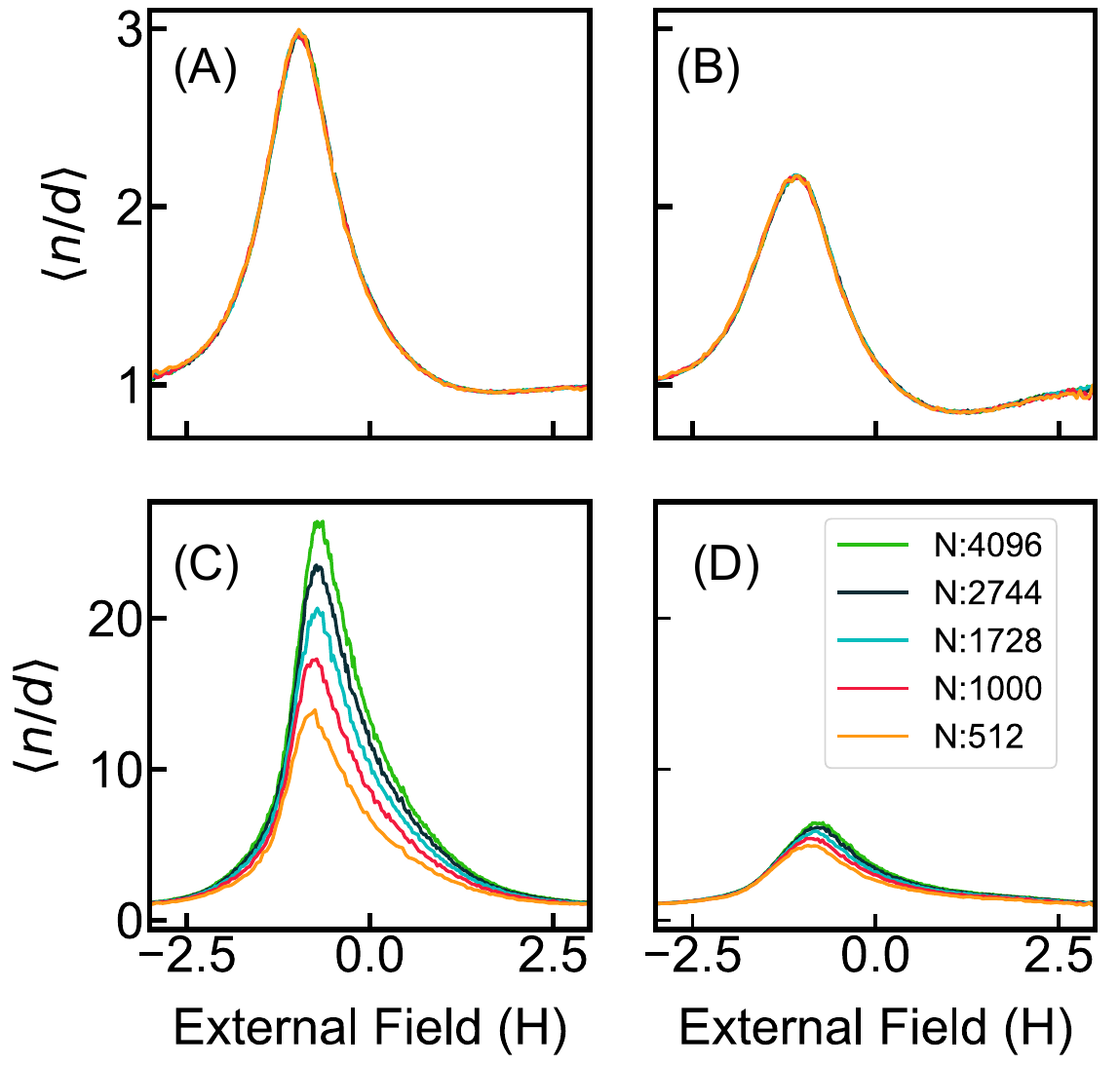}\hfill{}
\caption{\label{fig:impact} The ``impact" $ \langle n/d \rangle$ along the hysteresis loop, as a function of the external field $H$, is shown for fixed ramp $dH=c=0.25$ in panels (A) and (C) and for $dH=c=0.85$ in panels (B) and (D), respectively, for EA in the top row and SK in the bottom row. For EA, the ratio of the total number $n$ of spins that flip during an avalanche relative to the number $d$ of spins dislodged by the ramp $dH$ throughout the hysteresis loop remains independent of system size $N$. Even near an external field of $H \approx -1$, where the ramp $dH$ has the highest impact, each dislodged spin merely triggers at most $\approx2$ additional spins to flip when $c=0.25$, irrespective of system size. The impact of the ramp reduces further when $c$ is large, as shown in (B). For SK in panel (C), the impact diverges with system size, $n/d \rightarrow \infty$ for $N \rightarrow \infty$, which indicates criticality. However, in panel (D), SK begins to exhibit EA-like behavior when $c$ becomes large and the number of initially dislodged spins $d$ becomes so extensive as to limit the availability of other spins for subsequent flips.}
\end{figure}

For EA, avalanches along the hysteresis loop closer to saturated magnetized
states hardly occur, since there is rarely enough mutual frustration
in the system. Naturally, the largest response to a ramp can be expected
at the point where the susceptibility, $\chi=dm/dH$, is highest (usually
close to coercion, i.e., when $m=0$ at $H_{{\rm ext}}^{{\rm coer}}$,
see Fig.~\ref{fig:SKhyster}). However, conventional driving, i.e., mode
(1) for EA, at no point along the loop achieves avalanche sizes with
a correlation length anywhere close to system size, as panel (A) in
Fig.~\ref{fig:n-dependent} demonstrates. Yet, to facilitate a ``fair''
comparison with SK, EA would at least have to be driven in mode (2).
With the spacing between the most marginal stabilities $\lambda_{i}$
being $\sim1/N$, ramping with $dH\sim1/\sqrt{N}$ should dislodge
$\sim\sqrt{N}$ spins simultaneously throughout the lattice, while
SK in this mode merely triggers a finite number of spins. Still, in EA no
criticality emerges, since the correlation length is cut off independently and well before system-size effects emerge,
as panel (B) in Fig.~\ref{fig:n-dependent} shows. Merely a shift
in the overall duration $n$ of avalanches is observed that is commensurate
with the increase in the number of dislodged spins. We conclude
that asymptotically those $\sim\sqrt{N}$ small avalanches triggered
simultaneously throughout the lattice each remain too localized to
blend into larger correlated domains of flips that could percolate
the system. Thus, even for this side-by-side comparison of SK and
EA in mode (2), the conclusions of Refs. \citep{Goncalves08,Andresen2013}
remain applicable. 

Increasing the ramping rate to mode (3) for fixed $dH=c=0.25$, as
in panel (C) of Fig.~\ref{fig:n-dependent}, the distributions 
are broader but the form of $P(n)$ is system-size dependent, unlike in SK. 
In EA, a second peak emerges towards the end of the distribution, which becomes 
sharper with larger system sizes. Figure \ref{fig:c-dependent} shows that the same
effect occurs again when $c$ becomes larger as well, but now in both, EA and SK. 
These peaks are attributed to the number of spins dislodged ``simultaneously" upon 
the change in external field $H_{\rm ext}$. When the ramp rate is large enough, such 
as $c > 0.45$, far too many spin flips are triggered directly by the ramp of $dH$, overwriting any correlations needed to create critical behavior.  Even in SK, such spins include those that are not marginally 
stable, which suggests that marginally stable spins exclusively drive critical behavior.

The dependence of the secondary peaks on system size, as seen in EA, is investigated further in Figure \ref{fig:impact}. In order for either model to be critical, or create a broad cascade of flips, the total number of spins that flip, $n$, need to be much larger than the number of spins that were dislodged, $d$, due to the ramp. Fig.~\ref{fig:impact} therefore shows the average ratio $\langle n/d \rangle$, as a measure of the impact of a ramp. This value is measured along the changing external field $H$, since spin activity varies along the hysteresis loop. Panel (A) shows that $\langle n/d \rangle$ as a function of the varying external field remains invariant for changing system size. In EA, the number of spins that flip at all are therefore restricted by the number of spins that are dislodged. The distribution of $n/d$ itself (not shown) reveals that the skewness tends to zero with a growing system size. Evidently, the dominance of the dislodged spins in fixing the avalanche size becomes overwhelming for larger system sizes, hence the sharper peaks. Panel (B) shows the opposite effect for SK, suggesting that the impact of the dislodged spins here would asymptotically be infinite. Consistent with Fig.~\ref{fig:c-dependent}, SK begins to lose its critical behavior when the ``impact" of the dislodged spins weakens for large $c$, as shown in panel (D) and asymptotically becomes independent of system size. In this case, too many spins get dislodged by the ramp itself already, leaving few for any correlated activity.

\begin{figure*}
\hfill{}\includegraphics[width=160mm]{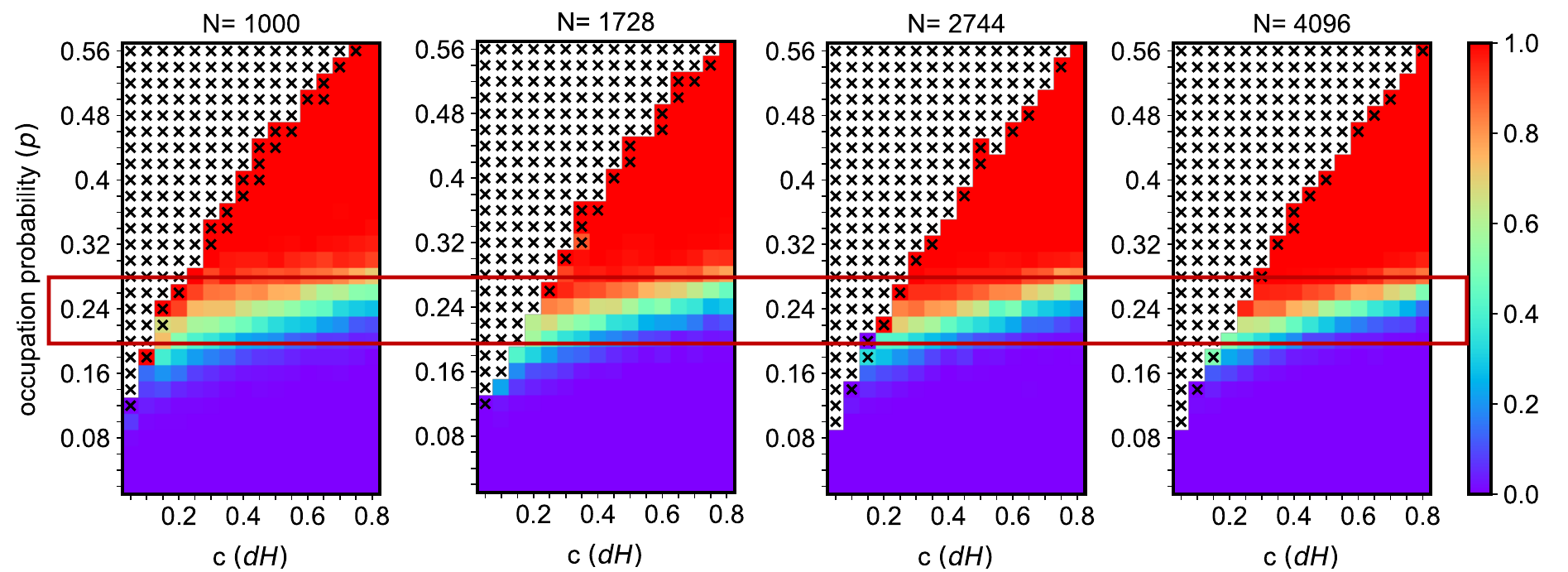}\hfill{}
\caption{\label{fig:percolation-grid} Each grid represents a different system
size $N$ for EA, and marks the probability of a percolating cluster, shown by the color bar, as a
function of both the constant ramping rate $c$, and the occupation
probability $p$, which is equivalent to the fraction of spins which
have flipped at least once during an avalanche. Note that each value
of $c$ produces a range of occupation probabilities, up to a maximum, mostly because
the distribution of spin flips changes for EA along the hysteresis
loop. Based on these statistics, the threshold at which there is an
onset of percolation emerges around $c\approx0.2-0.3$, which translates
into a critical occupation probability of $p_{c}\approx0.24$ (marked by red frame). The ``x" points mark probabilities that have been obtained with less than 100 data points. White spaces refer to combinations of $c$ and $p$ that have never occurred.}
\end{figure*}

\subsection{Spin Flip Avalanches as Percolation Clusters \label{subsec:Percolation}}

While any correlated, critical behavior in EA cannot be found in the statistics of the avalanche distributions, it can be expected that a correlated group of spins may form from one boundary of the cubic EA lattice to another to induce system-spanning avalanches for large enough values of a fixed ramp $dH=c$. This statistics only show cumulative effects during a given ramp, while avalanche sizes clearly vary from one ramp to the next. It is unclear whether correlated spin flips are conditioned on avalanches being a particular size. If we define the fraction of volume covered by all spins that flipped at least once during the avalanche triggered following a ramp as the occupation probability $p$, we might be able to describe any critical behavior in terms of a percolation transition for this mode of driving the hysteresis loop. 

In Fig.~\ref{fig:percolation-grid}, we study more closely the relation between the occupation probability $p$ generated by an avalanche for a given $c$ and the probability of that avalanche to percolate. Specifically, the color scale in Fig.~\ref{fig:percolation-grid} indicates the average percolation probability for an avalanche of given $p$ and $c$, i.e., the fraction  of avalanches in which percolation occurred, over the total number of avalanches across 5000 runs. It shows that for most values of $c$ there is a wide range of occupation probabilities $p$ and that the probability to percolate almost entirely depends on $p$ while it varies only weakly
with $c$. This range is consistent with the fact that we see avalanches
of different sizes along the hysteresis loop for the same $dH$. Certainly, the average occupation probabilities $\langle p\rangle$ attained rise with increasing $c$. However, more indicative of any transitional behavior is the maximum value of $p$ attained for a given $c$, marked by the upper boundary with the white space in Fig.~\ref{fig:percolation-grid}. For values near $c\approx0.2-0.3$ at every system size, occupation probabilities $p>0.24$ first appear that manage to produce spanning avalanches of a broad range of durations. Accordingly, we observe a transition from sub-critical to super-critical avalanches in the avalanche distribution $P(n)$ in Fig. \ref{fig:c-dependent} (B), the latter characterized by the emergence of a peak for the typical duration of a super-critical avalanche. 

In terms of the occupation probability $p$, we notice that spanning events arise quite close to the well-known threshold of $p_{c}\approx0.31160\ldots$ for ordinary site-percolation on a cubic lattice \citep{Grassberger92}. Spanning clusters at large values of $c$ are inevitable since many spins are forcibly flipped without any precondition of them being necessary correlated. However, for probabilities that are less than or near the random site percolation threshold $p_c$, potentially meaningful long-range correlations between flipping spins could emerge as previously disconnected clusters of activity could interact due to the induced spin flips. The indication of a threshold for spanning in this hysteretic avalanching process that is different from $p_{c}$ is a measure of how correlated frustration is among separated spins. Yet, as this threshold at $p\approx0.24$ appears to be just below $p_{c}$, and seems to approach $p_c$ further for
larger values of $N$ and $c$ in Fig.~\ref{fig:percolation-grid},
such correlations among spins remain weak even in this mode of hysteretic driving. The small difference between the random site percolation threshold $p_c$, and the experimental $p$ at which spanning clusters form, aligns with Fig.~\ref{fig:impact}. The extent of the short-range correlations is illustrated there -- a dislodged spin triggers only 2 additional spins to flip, and is therefore nearly independent. Aside
from some short-range correlations, on the larger scale, spins flip at almost random 
sites throughout the system. As the following discussion suggests, only a small amount of information 
is communicated among the flipping spins throughout those critical spanning clusters.

\begin{figure}
\hfill{}\includegraphics[width=0.85\columnwidth]{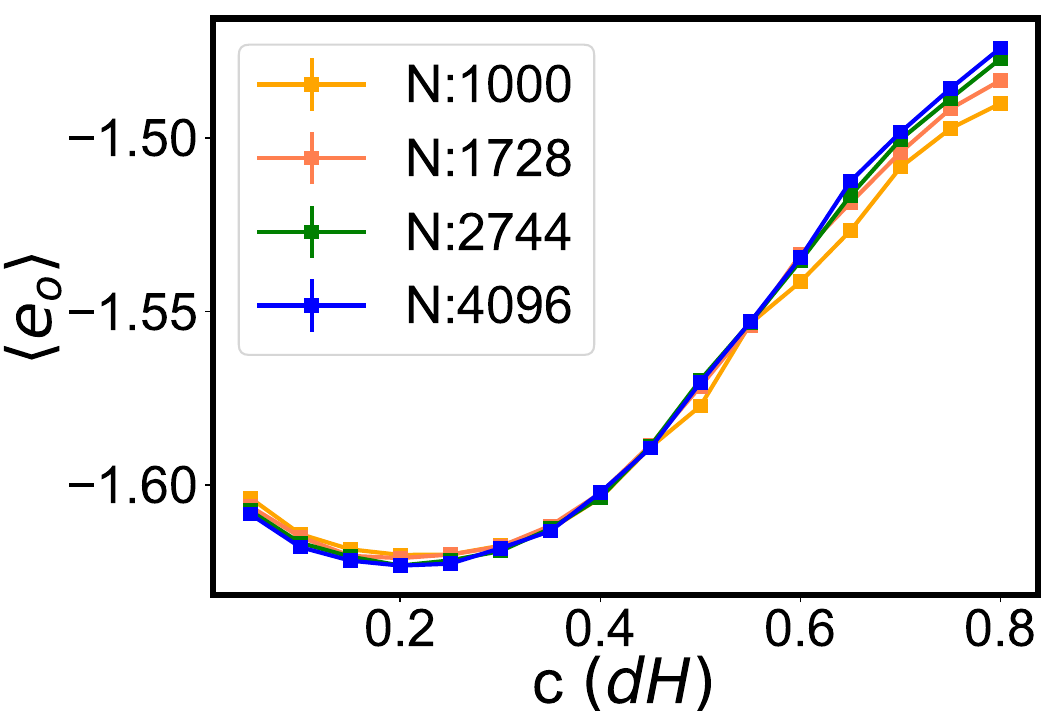}\hfill{}
\caption{\label{fig:average-ground-state} The approximate ground state energy
density $\langle e_0 \rangle$ for EA, averaged of 100 instances for each data point, 
as a function of $dH=c$, as reached in our modified HO heuristic. For given $N$, the same 
100 instances were used at each value of $c$. There is an alignment
between where the lowest approximation, reached at $c\approx0.2$, and where there is an onset
of percolation, see Fig.~\ref{fig:percolation-grid}. At extremely high c values, where percolation is frequent,
the large number of spins dislodged by the ramp overwrites all the helpful correlation that would have facilitated
cooperative behavior. If the value of $c$ is too low, the system is more
or less myopic and randomly chooses landscape exploration, whereas
a balanced value, $c\approx0.2$, ensures that the "right" spins are mutually
frustrated so that flipping them will encourage the most exploration. }
\end{figure}

\subsection{Percolating Spin Flip Avalanches and Optimization \label{subsec:Optimization}}

Aside from considerations of experimental feasibility, another motivation for understanding 
different modes of hysteretic driving is its
potential in optimization. Energy landscapes of disordered systems have many parallels
to NP-hard combinatorial optimization problems, requiring clever heuristics
to explore. Understanding the physics behind the driving mechanism that facilitates
long-range correlations, and perhaps a more global ordering,
in a disordered magnetic system can potentially help to design heuristics or adapt existing ones. 

For example, we could use the insights from the previous section to modify the hysteretic optimization heuristic (HO), amply described in Refs.~\citep{Pazmandi99,Pal06b,Goncalves08}, to adapt it to a sparse environment to better 
approximate ground states of EA. Originally, HO involves ramping the external magnetic 
field $H_{\rm ext}$ 
in the conventional manner (i.e., just flipping only the most unstable spin) back-and-forth to 
undergo hysteretic loops but of gradually diminishing widths. The expectation is that, 
when the width vanishes and the field converges to $H_{\rm ext}=0$, we arrive at a state 
with $m=0$ that is a good approximation to the ground state of the system.
Here, we replace the conventional ramps with fixed ones of size $dH=c$, independent 
of system size, to obtain a broader distribution of avalanches after each step.  
We start with an initial external field at $H_{\rm ext}^{0}=H_{\rm max}$, where $H_{\rm max}$ is just large enough so 
that the magnetization is saturated at $m=1$. The external field is then changed by $dH$ in the opposing direction, until $H_{\rm ext}^{t+1}=-\gamma H_{\rm ext}^{t}$ (one sweep, where $t$ denotes the sweep number, and $\gamma$ is a chosen parameter, which acts as a rate at which the hysteresis loop shrinks). This step is repeated in alternating directions between consecutive sweeps, until $H_{\rm ext}^{t}$ is close to zero -- to specify an exact value, we terminate the hysteresis once $\left| H_{t}\right|$ < $H_{\rm min}$, the minimum external field needed in order to trigger the least stable spin. 

Since the number of total steps along the hysteresis loop is dependent on $dH$, there is a chance that small $dH$ values will obtain lower energy states simply due to greater exposure. Accounting for this, the number of total sweeps for the annealing procedure is fixed to 1000, and the range of $\gamma$ is parameterized by $dH$. This shrinks the hysteretic loops faster for smaller $dH$ values, and slower for larger values, ensuring that all take roughly equal steps in total. To prevent the procedure from being deterministic, we designate a $\gamma_{\rm min}$ based on the geometric sequence $\gamma_{\rm min}  = 1 - \frac{2H_{\rm max}}{\eta c}$, where $\eta$ is the fixed number of sweeps. Then $\gamma$ is randomly selected from a uniform distribution between $\gamma_{min}$ and 1. This way, it becomes more likely for smaller $dH$ to have smaller $\gamma$, than is possible for larger $dH$.

With 100 instances of EA at system sizes $N=512$, 1000, 1728, 2744, and 4096, we
run this adapted version of the HO heuristic for a range of $0.05<c<0.8$ 
and obtained the best energy density, $\langle e_0 \rangle$ with $e_0={\cal H}_{\rm SG}/N$, 
using Eq.~(\ref{eq:SG-Hamiltonian}) rescaled by a factor of $\sqrt{2d}$, at $H_{\rm ext}=0$, averaged over those 100 instances 
for each value of $c$. The rescaling makes our experimental $\langle e_0 \rangle$ comparable to the known ``ground state" calculation of $\langle e_0 \rangle_{\rm opt}\approx1.70$ at $N\to\infty$~\citep{Pal96}, which uses $<J^2>=1$, unlike our EA model which uses Gaussian bonds with $<J^2>=1/2d$. In Fig.~\ref{fig:average-ground-state}, we plot the results for $\langle e_0 \rangle$ 
as a function of $c$ for each system size $N$. Aside from minute finite-size effects, 
$\langle e_0 \rangle$ consistently reaches a minimum near $c\approx0.2$. This estimate 
for the ground states of EA is better there than at any other value of $c$ or any other mode 
of driving the hysteresis loops. However, even this improved result is still quite far from any 
true optimum, \citep{Pal96}.   These findings are 
consistent with Sec.~\ref{subsec:Percolation}, where we observed that, for
a sparse system such as EA, only near the percolation threshold $p_c$ for spin flip clusters
do we have a chance of any long-range correlated behavior between spins that would allow 
them to explore their energy landscape to lower their frustration collaboratively. As Fig.~\ref{fig:percolation-grid}
shows, $c=0.2$ puts us close to that threshold.

\section{Conclusions}
\label{Conclusions}
We compare the effects of three different hysteretic driving modes in Ising spin glasses on both, an all-to-all connected model (SK) as well as a sparse lattice model (EA). Our study diverges from previous work for several reasons: First, we point out that conventional hysteretic driving protocols that trigger only the single most unstable spin leads not only to system-size dependence in the driving rate but also to inequitable driving rates between both models, since the average spacing between unstable spins scales with $\sim 1/N$ for EA (and any other sparse system), and $\sim 1/\sqrt{N}$ for SK. This implies that comparisons of the models, which have been made in previous literature, are inconsistent. However, side-by-side comparisons with equal driving protocols do not substantially change the previous conclusions. Second, we introduce an experimentally more realistic protocol of fixed ramps $dH=c$ that is equitable and system-size independent, in which (sub-)extensive sets of marginally stable spins are triggered to initiate each avalanche.
 Third, the exploration of a range of $c$-values allows us to study the role of marginally stable spins in inducing critical behavior. More specifically, we find that, to create critical fluctuations, the coordination of a prerequisite correlation structure is restricted only to marginally stable spins. Therefore, while it is possible for the SK model to lose its criticality by introducing ramps which interfere with marginal stability behavior, it is not possible to induce SK-like critical avalanches in EA, since no ramp is guaranteed to bring (only) marginally stable spins together.

\subsection*{Acknowledgements}
We thank Eric Weeks for his thoughtful suggestions and feedback during our conversations about this study.

\bibliographystyle{apsrev4-2}
\bibliography{/Users/sboettc/Boettcher}
 
\end{document}